# The "Golden Mode" at the Upgraded Tevatron?

S. Sultansoy

Deutsches Elektronen-Synchrotron DESY, Hamburg, Germany
Department of Physics, Faculty of Sciences and Arts, Gazi University, Ankara, Turkey
Institute of Physics, Academy of Sciences, Baku, Azerbaijan

**Abstract**

The existence of extra SM families results in essential enhancement of the gluon fusion channel Higgs boson production at hadron colliders. In this case, the SM Higgs boson can be seen at the upgraded Tevatron via the "golden mode" (H→4l) for certain values of its mass.

________________

Electronic address:    sultanov@mail.desy.de

It is known that the Standard Model (SM) does not predict the number of families of fundamental fermions. The LEP restriction $N_\nu = 3.00 \pm 0.06$ [1] is valid only for relatively light neutrinos with $m_\nu < m_Z/2$. On the other hand, the Flavor Democracy favors the existence of the nearly degenerate, heavy, $m_4 \approx 8 m_W$, fourth SM family (see [2] and references therein). Extra SM generations lead to essential enhancement of Higgs boson production due to the gluon fusion at hadron colliders [3-6], namely, $\approx 8$ times for four generations and $\approx 22$ times for five generations. With this enhancement gluon fusion mode becomes preferable for Higgs search at upgraded Tevatron. For example, as can be seen from analysis performed in [7] $gg \to H \to W^*W^*$ channel is very promising if 135 GeV $< m_H <$ 180 GeV. For lighter Higgs boson $gg \to H \to \tau^+\tau^-$ is the matter of interest [4]. For both cases the signal will be seen with more than $3\sigma$ statistical significance at $L^{int} > 10 fb^{-1}$.

In this note it is shown that extra generations will give opportunity to observe the "golden mode", $p\bar{p} \to HX \to l_1^+ l_1^- l_2^+ l_2^- X$ ($l_{1,2}=e,\mu$), at the upgraded Tevatron if 130 GeV $< m_H <$ 155 GeV or 185 GeV $< m_H <$ 250 GeV. Since the fifth SM family is strongly disfavored [2] by both precision electroweak data and phenomenology arguments, below I consider the four SM families. For estimations the value of $L^{int} = 15$ fb$^{-1}$ is used, which corresponds to the upgraded Tevatron run before the LHC will enter into game.

The expected numbers of "golden" events are given in Table 1. The Higgs production cross-sections are taken from [8]. Enhancement factor ($k_{4/3}$) values correspond to the degenerate fourth family with $m_4 \approx 640$ GeV [5]. It is seen that N(4l)>10 are predicted for two regions of the Higgs mass, namely, 130 GeV $< m_H <$ 155 GeV and 185 GeV $< m_H <$ 250 GeV. The most serious background is the pair production of Z bosons, $p\bar{p} \to ZZX$ ($Z \to l^+l^-$), which has $\sigma \approx 6$ fb [9] and should be taken into account for the latter region. This background can be suppressed by consideration of the appropriate four-lepton invariant mass distribution. For example, if $m_H = 200$ GeV, one expects 20 signal events, comparing to 5 background events within $M_{inv}(4l) = 200 \pm 5$ GeV bin.

So, the existence of the fourth SM family leads to essential consequences for Higgs search strategy at upgraded Tevatron (as well as at the LHC [6]). On the one hand, the Tevatron will be able to discover the Higgs boson if it is lighter than 250 GeV. On the other hand, if Higgs boson will not be found at the Tevatron and its mass is really less than 250 GeV, this will mean that the Nature really prefers the three SM families. Of course, these statements are based on rough estimations and detailed analysis of all three processes ($p\bar{p} \to HX$; $H \to \tau^+\tau^-$, $W^*W^*$ and 4l) and corresponding backgrounds should be performed. The work on the subject is under progress.

Finally, in the recent paper [10] it is shown that precision electroweak data allows the existence of a few extra SM families, if one allows new neutral leptons to have masses close to 50 GeV and sufficiently long lifetimes in order to escape the detector. In this case the Higgs boson becomes "invisible" at the LEP and Tevatron for $m_H <$ 160 GeV. Concerning the golden mode at the Tevatron, the first region (130 GeV $< m_H <$ 155 GeV) will be "lost", whereas our results are still valid for the second region (185 GeV $< m_H <$ 250 GeV). It seems that in the former case we should wait the TESLA [11] operation for the discovery of the Higgs boson. The work on this subject is also under progress.

P.S. In principle, similar enhancement takes place also in the framework of the MSSM at large tgβ [12]. However, I did not consider this possibility, since (in my opinion) SUSY should be realized at (pre-)preonic level, because of the parameter inflation (see [13] and references therein).

**Acknowledgements**

I am grateful to I. Ginzburg, L. Gladilin and A. Turcot for useful discussions and valuable remarks. I would like also to express my gratitude to DESY Directorate for invitation and hospitality.

**References**

1. D.E. Groom et al., Eur. Phys. J. C **15**, 1 (2000).
2. S. Sultansoy, *Why the Four SM Families*, contributed paper to ICHEP2000; hep-ph/0004271.
3. V.D. Barger and R.J.N. Philips, Collider Physics, Addison-Wisley, 1997.
4. I.F. Ginzburg, I.P. Ivanov and A. Schiller, Phys. Rev. D **60** (1999) 095001.
5. E. Arik et al., ATLAS Internal Note ATL-PHYS-98-125 (1999).
6. ATLAS Detector and Physics Performance Technical Design Report, CERN/LHCC/99-15 (1999), p. 663.
7. T. Han, A.S. Turcot and R.-J. Zhang, Phys. Rev. D **59**, 093001 (1999).
8. Run II Higgs Working Group of the Run II SUSY/Higgs workshop: http://fnth37.fnal.gov/higgs.html.
9. J. Ohnemus, Phys. Rev. D **50** (1994) 1931.
10. M. Maltoni et al., Phys. Lett. B **476** (2000) 107.
11. R. Brinkmann, G. Materlik, J. Rossbach and A. Wagner (Eds), DESY 1997-048/ECFA 1997-182 (1997).
12. M. Spira, DESY Preprint 98-159; hep-ph/9810289.
13. S. Sultansoy, Ankara University Preprint AU-HEP-00-01; hep-ph/0003269.

Table 1. The expected numbers of golden events in the four SM families case at 15 fb$^{-1}$.

| $M_H$, GeV | $\sigma_3(p\bar{p} \to HX)$, pb | $k_{4/3}$ | $BR(H \to ZZ)$, $10^{-2}$ | N(4l) |
|---|---|---|---|---|
| 120 | 0,.70 | 8.7 | 1.5 | 6 |
| 130 | 0.56 | 8.6 | 3.8 | **12** |
| 140 | 0.45 | 8.5 | 6.7 | **17** |
| 150 | 0.36 | 8.4 | 8.2 | **17** |
| 160 | 0.30 | 9.3 | 4.1 | 7 |
| 170 | 0.25 | 8.2 | 2.2 | 3 |
| 180 | 0.20 | 8.1 | 5.7 | 6 |
| 190 | 0.17 | 8.0 | 22 | **20** |
| 200 | 0.14 | 8.0 | 26 | **20** |
| 210 | 0.12 | 7.9 | 29 | **19** |
| 230 | 0.09 | 7.7 | 31 | **14** |
| 250 | 0.07 | 7.5 | 32 | **11** |